\documentclass[sigconf]{acmart}
\AtBeginDocument{%
  \providecommand\BibTeX{{%
    \normalfont B\kern-0.5em{\scshape i\kern-0.25em b}\kern-0.8em\TeX}}}

\setcopyright{acmlicensed}
\copyrightyear{2018}
\acmYear{2018}
\acmDOI{XXXXXXX.XXXXXXX}
\usepackage{tabularx}
\usepackage{mdframed}

\acmConference[Conference acronym 'XX]{Make sure to enter the correct
  conference title from your rights confirmation email}{June 03--05,
  2018}{Woodstock, NY}
\acmBooktitle{Proceedings of the 32nd ACM Symposium on the Foundations of Software Engineering (FSE '24), November 15--19, 2024, Porto de Galinhas, Brazil}

\begin{document}

\title{The Role of Generative AI in Software Development Productivity: A Pilot Case Study}

\author{Mariana Coutinho}
\affiliation{
  \institution{CESAR School}
  \city{Recife}
  \state{PE}
  \country{Brazil}
}
\email{mclc@cesar.school}

\author{Lorena Marques}
\affiliation{
  \institution{CESAR School}
  \city{Recife}
  \state{PE}
  \country{Brazil}
}
\email{lmvs@cesar.school}

\author{Anderson Santos}
\affiliation{
  \institution{CESAR}
  \city{Recife}
  \state{PE}
  \country{Brazil}
}
\email{acss@cesar.org.br}

\author{Marcio Dahia}
\affiliation{
  \institution{CESAR}
  \city{Recife}
  \state{PE}
  \country{Brazil}
}
\email{mlmd@cesar.org.br}

\author{Cesar França}
\affiliation{
  \institution{CESAR}
  \city{Recife}
  \state{PE}
  \country{Brazil}
}
\email{franssa@cesar.org.br}

\author{Ronnie de Souza Santos}
\affiliation{%
  \institution{University of Calgary}
  \city{Calgary}
  \state{AB}
  \country{Canada}
}
\email{ronnie.souzasantos@ucalgary.ca}

\copyrightyear{2024}
\acmYear{2024}
\setcopyright{acmlicensed}\acmConference[AIware '24]{Proceedings of the 1st
ACM International Conference on AI-Powered Software}{July 15--16,
2024}{Porto de Galinhas, Brazil}
\acmBooktitle{Proceedings of the 1st ACM International Conference on
AI-Powered Software (AIware '24), July 15--16, 2024, Porto de Galinhas, Brazil}
\acmDOI{10.1145/3664646.3664773}
\acmISBN{979-8-4007-0685-1/24/07}

\begin{abstract}
With software development increasingly reliant on innovative technologies, there is a growing interest in exploring the potential of generative AI tools to streamline processes and enhance productivity. In this scenario, this paper investigates the integration of generative AI tools within software development, focusing on understanding their uses, benefits, and challenges to software professionals, in particular, looking at aspects of productivity. Through a pilot case study involving software practitioners working in different roles, we gathered valuable experiences on the integration of generative AI tools into their daily work routines. Our findings reveal a generally positive perception of these tools in individual productivity while also highlighting the need to address identified limitations. Overall, our research sets the stage for further exploration into the evolving landscape of software development practices with the integration of generative AI tools.
\end{abstract}



\maketitle

\section{Introduction}
Over the last decade, software engineering research has explored various aspects of teamwork in software development. Such investigations involved understanding how software engineers work together, how they collaborate to tackle problems, how they share information to complete tasks, their choices in adopting tools, and the obstacles they encounter in their activities \cite{prikladnicki2013cooperative, strode2022teamwork, hoegl2003teamwork}. Understanding these aspects is crucial for improving individual and team performance and ultimately achieving success in software processes. Despite these research efforts, understanding productivity in software development remains challenging due to its multifaceted nature \cite{hoegl2003teamwork, lindsjorn2016teamwork}. Productivity is influenced by technical, social, and psychological factors, making it complex to fully grasp. Additionally, subjective metrics, diverse tasks, and evolving team dynamics add further layers of complexity \cite{rodriguez2012empirical, guerrero2022team}. Therefore, despite progress in understanding teamwork in software engineering, unlocking the essence of productivity in software teams remains difficult \cite{guerrero2022team, sadowski2019rethinking, forsgren2021space}.

Currently, the introduction of generative AI has elevated investigations into productivity in software development to a new level as discussions shift towards the incorporation of generative AI-based tools to enhance productivity, increase work efficiency, reduce errors in software tasks, and accelerate software production \cite{noy2023experimental, ebert2023generative, nam2024using}. Notably, tools like GitHub Copilot have emerged as promising coding aids to improve code writing time. However, despite the general belief in the potential benefits that generative AI brings to software development, particularly regarding productivity gains, empirical evidence remains scarce, especially when considering the context of complex real-world projects, such as industrial settings \cite{peng2023impact, monteiro2023end}.

The lack of empirical evidence regarding the effectiveness of generative AI in complex real-world software development projects has motivated the present research. Hence, our goal was to explore how generative AI tools might impact the productivity of software professionals working on different roles and activities, including those focused on the delivery of software (e.g., coding and testing), supporting activities (e.g., management and IT infrastructure) and related tasks (e.g., data science). To this end, we investigated this phenomenon in a large software company. Specifically, the following research question guided this study: \textit{\textbf{How does the integration of generative AI tools influence the work of software professionals across different roles and activities?}} We are particularly interested in investigating the relationship between the usage of these tools and aspects associated with productivity.

The remainder of this paper is organized as follows. In Section~\ref{sec:back}, we present a literature review on productivity and generative AI in software development. In Section~\ref{sec:method}, we present the case under study, our data collection, and our data analysis strategy. In Section~\ref{sec:findings}, we introduce the key insights derived from the case, which are discussed in Section~\ref{sec:discussions}. Finally, Section~\ref{sec:limit} discusses the limitations of our study, and Section~\ref{sec:conclusions} outlines our conclusions and plans for future work.

\section{Background} \label{sec:back}
In this section, we explore previous research on productivity and generative AI in software development, specifically addressing the challenges associated with measuring productivity and integrating generative AI tools into software development tasks. 

\subsection{Challenges in Productivity Analysis} \label{sec:productivity}
Understanding and measuring productivity among software developers has long been a challenge due to the absence of universally accepted metrics. The complex nature of development tasks, coupled with multifaceted dimensions of productivity, poses significant hurdles in devising accurate metrics. Factors such as individual work styles, team dynamics, task complexities, and subjective perceptions contribute to the intricate web of challenges inherent in measuring developer productivity \cite{forsgren2021space, guerrero2022team}. 

Diverse industry discussions commonly refer to productivity as the relationship between output and input, but diverse fields adopt varying notions and measurement units ~\cite{sadowski2019rethinking}. However, looking specifically at software engineering, these conventional metrics, such as considering lines of code as the relationship between input and output, inadequately capture the essence of software development since this work thrives on collaborative efforts stemming from a diverse group of individuals, each contributing unique expertise \cite{forsgren2021space}. 

Software developers operate within the domain of knowledge workers, where productivity lies in the exchange of ideas, knowledge, and skills, where collaboration becomes the cornerstone, fostering innovation and comprehensive solutions. ~\cite{forsgren2021space, ruvimova2022exploratory}. These encompass individual values, professional objectives, and the standards established by the organization. Moreover, their work revolves around activities that demand creativity as part of their responsibilities ~\cite{kim2019understanding}.

Historically, the evolution of productivity measures and discussions in software development reflects a quest for a comprehensive definition that considers multifaceted aspects influencing project outcomes \cite{albrecht1979}. Over the years, diverse methodologies emerged, from quantifying delivered code to embracing Agile principles, and the rapid evolution in technology continues to shape this quest, as changes in software practices and tools directly impact the productivity of software professionals \cite{albrecht1979, boehm2000, wohlin1995soft, lakhanpal1993}. For instance, generative AI has recently emerged as a prominent topic, influencing the debate on productivity in software development.

Recently, expanding on several research insights, important dimensions of developers' productivity were highlighted, including ~\cite{forsgren2021space}: 

\begin{itemize}
    \item \textbf{Satisfaction and well-being}: Focuses on happiness, job satisfaction, work-life balance, feeling valued, and a positive work environment. High levels of satisfaction boost productivity, creativity, and retention.
    
    \item \textbf{Performance}: Encompasses delivering high-quality work efficiently and accurately within set timelines. Metrics like lead time for changes, deployment frequency, and change failure rate gauge performance, impacting customer satisfaction and business outcomes.
   
    \item \textbf{Activity}: Evaluates the volume and nature of teamwork, such as completed tasks, lines of code, and commits. However, activity metrics shouldn't be the sole focus as they might not directly correlate with productivity or value delivered.
    
    \item \textbf{Communication and collaboration}: Assesses how well team members communicate, share knowledge, and work together towards common goals, vital for successful software development teamwork.
    
    \item \textbf{Efficiency and flow}: Efficiency involves optimizing workflows to maximize output by eliminating bottlenecks and unnecessary steps. Flow describes a state of full immersion, fostering creativity, focus, and efficiency among team members.
\end{itemize}

These dimensions challenge myths and misconceptions surrounding developers' productivity, demonstrating that only the combination of several metrics allows for a nuanced understanding of productivity in software engineering.

\subsection{Productivity and Generative AI} \label{sec:genAI}
Recent studies have shown that generative AI can enhance productivity in software development primarily through automated code generation \cite{li2024sheetcopilot}. In this context, large language models are trained on several code datasets to produce usable code in response to specific task prompts. While common chat interfaces can generate both code and natural language responses, tools designed specifically for code generation, such as GitHub Copilot, CodeWhisperer, and ChatGPT, have gained popularity among developers \cite{cambon2023early, sikand2024much}.

GitHub Copilot, developed collaboratively by GitHub, OpenAI, and Microsoft, is described as an AI pair programmer, providing developers with real-time code suggestions based on the context of comments and existing code, supporting their productivity by minimizing disruptions and increasing focus on several aspects of programming \cite{githubcopilot}. Additionally, CodeWhisperer is an AI code generator by AWS that offers real-time code suggestions as developers write, anticipate the completion of lines of code, comments, or generate entire functions and code blocks, allowing faster task completion \cite{aws_codewhisperer}. ChatGPT, developed by OpenAI, serves as a conversational AI and virtual assistant, helping developers in coding, debugging, and learning, ultimately enhancing productivity in software development \cite{OpenAIChatGPT}.

Preliminary studies focused on the use of these tools and discussed their potential effects on the productivity of developers. In ~\cite{peng2023impact}, the authors highlighted the positive correlation between the use of GitHub Copilot and the productivity of developers, with a focus on the automation of repetitive tasks. Following this, ~\cite{github_copilot_impact} expands this scenario and confirms the positive correlations while also recognizing the need for broader productivity metrics beyond coding time to obtain insights on developer satisfaction. Further, ~\cite{mckinsey_generative_ai} also explores the speed gains offered by generative AI-based tools, acknowledging that these gains vary with task complexity and developer experience. Lastly, ~\cite{zhang2023practices} offers a comprehensive review of AI integration trends in software development, including the increased role of AI in task automation and decision support.

The reports primarily focus on task completion time as a productivity metric, but productivity in software development entails complex factors beyond this \cite{sadowski2019rethinking}. In \cite{monteiro2023end}, a more comprehensive investigation details an application of ChatGPT in software development, covering the entire software creation process from requirements to deployment. Though productivity measurement is not the primary focus, the study offers insights into the nuanced challenges of measuring the effects of generative AI on productivity.

\section{Method} \label{sec:method}
Recognizing the limited empirical evidence from real-world settings regarding the use of generative AI to enhance productivity in software development, we chose a methodology centered on a case study \cite{yin1994discovering}. Software engineering case studies analyze real-life settings, like companies or teams, using a systematic approach to collect and analyze data that can inform industrial practice \cite{runeson2009guidelines}. In this research, prior to conducting a comprehensive investigation considering the nuanced nature of productivity, we began with a simplified pilot case study \cite{xie2008using, monteiro2016innovative}. This pilot study aimed to gather insights from developers who were using these tools for the first time as an official working tool. By focusing on this initial experience from developers, we plan to gain a preliminary understanding of how we can explore the nuanced characteristics of productivity in this context. Further details regarding our methodology are outlined in the sections below.

\subsection{The Case}
The company selected for our case study was established in 1996 and specialized in on-demand software solutions across various sectors like finance, telecommunications, government, manufacturing, services, and utilities. With a workforce exceeding 1,200 professionals, over 70\% are directly engaged in software development across 50 distinct teams. These teams comprise individuals from diverse technical backgrounds, including programmers, quality assurance (QA) specialists, and designers, who are proficient in popular software development methodologies such as Scrum, Kanban, and Waterfall and work to develop systems for global clients spanning North America, Latin America, Europe, and Asia. Additionally, the company employs several professionals in related fields like data science and supporting functions such as IT and human resources.

This company provides an excellent setting for our case study as it offers diverse tasks ranging from common coding activities to other crucial stages of the software development process. By exploring the impact of generative AI tools across these tasks, we can gain comprehensive insights. Moreover, given that various contextual factors and backgrounds influence productivity, the company's involvement in multiple industrial sectors allows us to assess productivity within varied contexts, enhancing the study's robustness and applicability. Additionally, the company's interest in experimenting with the use of generative AI tools provides an opportunity to leverage the insights gained towards devising processes and policies for their integration into projects.

\subsection{Defining the Pilot Study Scope}

Given the diverse array of projects, contexts, backgrounds, and professionals within the company, we chose to refine our research approach by initiating a preliminary investigation within a more focused subset. This entailed selecting a specific group of individuals to engage with AI generative tools and document their experiences. By focusing on this targeted group, we aimed to gather valuable insights that would guide us in the formulation of a broader, more extensive investigation within the case, exploring various productivity facets in software development.

During this initial phase, we obtained 17 licenses for various tools, including ChatGPT Plus, OpenAI API, Midjourney, and GitHub Copilot, for professionals to utilize in their work and provide feedback on these tools. In this pilot study, any interested professional could participate, provided they met three specific criteria: a) they hadn't regularly used generative AI tools in their work (priority given to those who hadn't used them at all); b) they received approval from their team manager to integrate the tool into their tasks; c) they committed to reporting their experiences, including both positive and negative impacts on their perceived productivity.

After extending invitations to participate via the company's communication channels, we selected 14 volunteers who were either actively engaged in software development or working in supporting roles to join the pilot case study. These participants were chosen based on the aforementioned criteria, with diversity across backgrounds, experience levels, and project types considered to ensure broad representation. This approach aimed to capture a spectrum of perspectives and experiences, thereby enhancing the insights derived from the pilot study.

\subsection{Data Collection}
In line with the case study methodology, we employed various data collection methods to explore the case, including questionnaires with open-ended questions \cite{molleri2016survey} and observations \cite{seaman1999qualitative}. The primary data collection technique utilized was the questionnaire, which directly gathered insights from the professionals who volunteered to use the AI tools. Unlike traditional case studies that often rely on interviews, we opted for a questionnaire-based approach in this pilot study. This decision was made because the use of the tools was voluntary and not consistently integrated into the volunteers' daily work life or projects. We aimed to minimize interference with their work dynamics or team activities. Additional data were obtained from observing the company's communication channels, such as Slack, both to identify potential study volunteers and to explore what other professionals in the company were discussing about generative AI. This data collection process spanned four weeks during the final months of 2023.

The questionnaire (Table \ref{tab:questionnaire}) was crafted to elicit descriptive answers, focusing on qualitative data from the study participants. Questions were designed to capture various aspects of their experiences with the provided AI tools, shedding light on the benefits, challenges, and overall impact on productivity and workload. For instance, participants were prompted to summarize their activities and experiences with the AI tool, identify the main benefits and difficulties encountered, and reflect on the tool's influence on productivity and value creation in their tasks. Moreover, the questions explored the participants' satisfaction with the results and their willingness to continue using the AI tool. Through these questions, we aimed to gather insights into the participants' experiences and perceptions regarding the use of generative AI tools in software development that could support us in the establishment of a comprehensive longitudinal case study.


\begin{table}
\caption{Survey Questionnaire}
\label{tab:questionnaire}
\scriptsize
\begin{tabularx}{\linewidth}{p{8cm} X }\midrule
1. This form is designed to gather feedback from team members who have been granted licenses to utilize Generative AI tools. The data collected will contribute to an Experience Report paper focusing on how Generative AI can enhance organizational productivity. Rest assured, the information provided will be handled with utmost confidentiality and security. Any materials derived from this study will utilize anonymized data, ensuring the privacy of respondents. No personal or sensitive information will be disclosed to third parties without prior consent.
Do you agree to participate?  \\
( ) Yes \\ \midrule

2. Which tool did you receive a license for? \\
( ) ChatGPT \\
( ) Open AI API \\
( ) Midjourney \\
( ) GitHub Copilot \\ \\

3. What is your role in your team? \\ \\ 

4. Please give a brief summary of the activities you performed with the provided AI tool. \\ \\ 

5. Share details of your experience of working with the provided tools. \\ \\ 

6. Explain the main benefits you encountered while using the provided AI tool.\\\\ 

7. Describe any impact you observe on your productivity while using any of the tools.\\ \\ 

8. Did the tool contribute to enhancing the value of your activities? For instance, did it help you make progress in tasks or provide other valuable insights? If yes, how did it contribute?\\ \\ 

9. You were satisfied with the results of the tasks performed with the assistance of generative AI?\\ \\ 

10. Would you like to continue using AI in your work? \\ \\ 

11. Describe any other important or interesting aspects of your experience. \\ 

\bottomrule
\end{tabularx}
\end{table}

\subsection{Data Analysis}
Our data analysis process involved examining responses to open-ended questions in the questionnaire and notes from observing interactions in communication channels. We employed coding strategies, including line-by-line open \cite{charmaz2014constructing} supplemented by thematic analysis \cite{cruzes2011recommended}. Through this approach, we identified recurring themes and patterns related to participants' experiences with AI tools, including benefits, challenges, productivity impact, satisfaction levels, and willingness to continue using generative AI tools in software development.

\section{Findings} \label{sec:findings}
Initially, we had 14 individuals participating in our pilot study. However, one individual withdrew from the study for not being able to incorporate the tool usage in their daily activities considering the current phase of their project, resulting in a final group of 13 software professionals with various roles within software teams. It includes four software engineers, four software designers, three data scientists engaged in software projects, one software QA specialist, and one agile coach. All members actively participate in software development activities across different projects. Furthermore, our participants comprise an experienced cohort of professionals, with seven at a mid-level, two senior professionals, two principal professionals, and one technical manager, demonstrating a breadth of expertise and leadership within the field. Below, we present the findings derived from their experiences with the AI tools integrated into their regular activities within their projects. This encompasses insights into their utilization patterns, identified benefits, encountered challenges, and perceived impacts on productivity. In Table \ref{tab:findings}, we present evidence extracted from the participants' narratives that support our findings. 

\subsection{Generative AI Tools: Uses} \label{sec:uses}
The software professionals participating in our pilot study have utilized generative AI tools with a common objective: to acquire new knowledge across various aspects of their work. Among the cited uses, we identified:

\begin{itemize}

\item Generating and Reviewing Artifacts: Involves using AI tools to review, refine, and produce various project documents, such as requirements specifications, design documents, or project plans, ensuring accuracy and completeness in project deliverables.

\item Supporting Ideation Processes: Entails harnessing AI to support the development of novel ideas, concepts, or solutions, aiding, for instance, in brainstorming sessions and design thinking.

\item Resolving Doubts in Code Construction: Involves leveraging AI to assist in resolving technical issues encountered during programming activities.

\item Conducting Formal Writing: Refers to utilizing AI-powered natural language generation tools to aid in the creation of formal project documentation, such as project reports, technical documentation, or client presentations.
\end{itemize}

Considering these findings, we can observe that the overarching goal in using generative AI in software activities remained consistent—using acquired knowledge to optimize work processes and facilitate problem-solving. Additionally, AI tools have played an interesting role in improving communication among stakeholders. These activities underscore a common thread of utilizing AI to streamline workflows and improve team capabilities across diverse tasks and domains.

\subsection{Generative AI Tools: Benefits} \label{sec:benefits}
In our analysis, we identified two main benefits highlighted by the participants, namely, time optimization and versatility. Participants in the study consistently highlighted the tool's support for time optimization as a primary benefit. By harnessing generative AI, software professionals were able to save time while completing their tasks. In particular, this time saving was apparent through the support for various activities that involve writing artifacts, such as reports. The tool's ability to generate coherent and relevant content and provide valuable insights and suggestions significantly supported participants' writing activities, enabling them to produce these types of artifacts with greater ease. Furthermore, participants highlighted the AI tools' versatility in supporting a wide range of software tasks as a visible benefit. The effectiveness of providing timely and relevant support across these diverse tasks was demonstrated to be a valuable advantage of the tools.

\subsection{Generative AI Tools: Challenges} \label{sec:challenges}
The participants reported challenges associated with utilizing generative AI in their software development, with reliability and refinement emerging as the most recurring issues. Software professionals described encountering difficulties in ensuring the reliability and functionality of generated responses, especially when attempting to use multiple questions simultaneously. They also faced challenges in crafting precise prompts to obtain objective and accurate responses, along with concerns about the absence of sources to reinforce the reliability of results. Additionally, participants highlighted the need for refinement and fine-tuning of the generated results to achieve optimal usage. Despite the AI's capability to produce responses, participants found that the outputs often required manual adjustments and polishing before they could be effectively incorporated into their work. Finally, security measures emerged as a potential challenge, with at least two participants noting their inability to use the support of the tools with sensitive project data due to security constraints.

\subsection{Generative AI Tools: Effects of Perceived Productivity} \label{sec:productivity}
Aligned with the previous findings, participants reported a positive effect of generative AI tools on their perceived productivity. They highlighted how these tools facilitated efficiency gains in various aspects of their software development activities, thus primarily relating the optimization of time with productivity gains. One notable characteristic of this optimization was the consolidation of several individual tools into a single tool to perform various activities. Therefore, despite encountering challenges such as reliability concerns or limited outcomes, with the exception of one individual, software professionals mostly reported a positive impact on their perceived productivity.

On an additional note, software professionals related their increased productivity with the value that the AI tools incorporated into their work, especially in facilitating the creation of relevant and insightful content, whether reports, code, or design models. More specifically, the tools supported their productivity by contributing to learning and knowledge acquisition by providing quick access to information because, despite needing external verification, using these tools is much more productive than seeking information through queries in search engines.

\begin{table*}
  \caption{Evidence Obtained from Participants}
  \label{tab:findings}

\begin{tabularx}{\linewidth}{p{1.8cm} p{2.2cm} X}
\toprule
Finding & Theme & Papers \\ \hline 

Usage & Generating and Reviewing \newline Artifacts & ``The primary activity I engaged in with the tool was revisiting the questions from the TDD form.'' (P01) \newline ``I used the tool to support desk research activities, to create questions for interviews.'' (P06) \newline ``I used it to conduct a proof of concept for audio transcription.''(P12) \newline \\

 & Supporting Ideation \newline Processes & ``to experiment with ideation processes. I also used it to support the creation of presentation material''. (P06) \newline ``I used midjourney as part of the creative process, combining prompts and images we already use to generate new texture ideas.'' (P07) \newline\\ 
 
 & Resolving Doubts in Code Construction & ``I used the tool to solve some small doubts about Python code construction.'' (P01) \newline `` I was able to evaluate how the tool can support data exploration processes, code generation, and others.'' (P03) \newline ``The tool was used for code syntax research, automatic generation of simple algorithms, writing unit tests.'' (P11) \newline \\

 & Conducting \newline Formal Writing & ``The use of ChatGPT has been very useful at the beginning of text production activities.'' (P04) \newline ``Aiding in composing texts from topics, suggesting ideas for product and process names.'' (P09) \newline ``I have been using the OpenAI GPT-4 model for a variety of tasks, including text generation.'' (P10)\\
\hline

Benefits & Time \newline Optimization & ``It allowed me to be faster in text writing situations and debugging.'' (P09) \newline ``The tool has proven to be useful in saving time and effort, allowing me to focus on more critical tasks.'' (P10) \newline ``The main gain was in development and research time.'' (P12) \newline \\

& Knowledge \newline Acquisition & ``I believe the main benefit was what I learned.'' (P03) \newline ``Initially, I sought information I was already familiar with to validate my knowledge.'' (P05) \newline ``It is a very powerful ally in consuming information, verifying and summarizing theories and approaches.'' (P13) \newline  \\

& Versatility & ``I believe that using our own materials as data and making the AI generate new combinations.'' (P07) \newline ``The flexibility and adaptability of the tool have been crucial aspects that enhance its value and applicability in various contexts..'' (P10) \newline ``It is a much more productive way of seeking knowledge when compared to the model of using search engines (e.g., Google) up to that point.'' (P13)  \\
\hline

Challenges & Reliability & ``Difficulty in verifying the reliability of some information.'' (P06) \newline ``I faced challenges, mainly in ensuring that the generated responses are accurate and reliable, which sometimes requires manual review and adjustment.'' (P10) \newline ``The absence of sources is one of the main barriers to the reliability of the results.'' (P13) \newline \\
& Precision & ``Continuous use, makes the results less interesting.'' (P04) \newline ``The main difficulties were in achieving more 'usable' results; all items generated still need refinement.'' (P07) \newline ``Writing quality prompts for objective and correct answers.'' (P09) \newline \\
& Security & ``The adoption of a model that I can use sensitive data from the company.'' (P04) \newline ``I would like to add a crucial observation about the ethical and privacy challenges associated with the use of AI tools like GPT.'' (P10) \newline ``The main difficulty was not exposing code used in clients.'' (P11) \\
\hline

Productivity &  & ``
Absolutely. The value in the productivity and speed of generating results from my prompts is undeniable..'' (P05) \newline ``Yes, the AI tool provided significant value in various areas of my activities. Firstly, it significantly improved my efficiency.'' (P10) \\ 

\bottomrule
\end{tabularx}

\end{table*}

\section{Discussions} 
\label{sec:discussions}
We focused our discussions on the nature of software developers' productivity presented in the literature, particularly emphasizing that software developers operate within the domain of knowledge workers, where productivity lies in the exchange of ideas, knowledge, and skills. In this context, we compared how the positive impact of using generative AI tools reported by participants aligns with several dimensions of productivity, namely, satisfaction and well-being, performance, activity, communication and collaboration, and efficiency and flow.

Primarily, participants emphasized how these tools improve \textit{efficiency and flow}, with gains being observed in various aspects of their software development activities. By optimizing time and consolidating multiple tools into a streamlined workflow, participants maximized the efficiency of their outputs with less effort. Moreover, the reported positive impact on productivity suggests that participants experienced an improvement in their \textit{performance} as they were able to create relevant and insightful content, such as reports, code, or design models.

Additionally, while not explicitly mentioned by the participants, we understand that the use of generative AI tools can indirectly impact \textit{communication and collaboration} within software development teams. By providing quick access to information and facilitating knowledge acquisition, these tools can enhance communication and collaboration by enabling team members to share insights and align their understanding toward common goals.

\subsection{Implications} 
Our study has implications for research, as our pilot case study takes on a real-world perspective to explore the evolving landscape of software development practices with the integration of generative AI tools. As participants expressed positive experiences despite encountering challenges, they suggested that the benefits of these tools outweighed the drawbacks. Even though our findings are preliminary, they underscore the need for further investigation into the efficacy and impact of these tools across various dimensions of productivity. In particular, we highlight the need for research focused on refining these tools to address reliability concerns and expand their capabilities, particularly around the dimensions of productivity not identified in this study, e.g., activity and satisfaction.

Additionally, our study has implications for industrial practice, as it sheds light on the practical benefits of integrating generative AI tools into software development workflows, considering different professional roles, including programming, testing, and design. The positive experiences reported by participants indicate the potential of these tools to enhance productivity. Therefore, by addressing challenges and leveraging the advantages offered by generative AI, software companies can potentially optimize their development processes. Moreover, considering the challenges reported by practitioners regarding reliability concerns and usage difficulties, our findings suggest the importance of providing adequate training and support to facilitate the effective adoption of these tools within development teams, ensuring that they maximize their potential benefits while minimizing any associated risks.

\subsection{Future Work} 
Following our pilot case study, our immediate future work involves conducting a comprehensive case study, capitalizing on the availability of the company that participated in this study. Our focus will be twofold. Firstly, we aim to explore the particularities arising from the utilization of generative AI tools across various software development roles, ranging from developers to QAs and designers, thereby gaining insights into how different professionals perceive and utilize these tools within their specific tasks. Secondly, we aim to further explore the relationship between generative AI tools and the dimensions of productivity by expanding our participant cohort within the case study to encompass a diverse range of project configurations and software development methodologies. By doing so, we aim to offer a more detailed analysis of the impact of generative AI tools on productivity across different software development contexts, thereby facilitating a more nuanced discussion on this subject.

\section{Threats to Validity} \label{sec:limit}
While our pilot case study provides valuable insights into integrating generative AI tools into software development workflows, some limitations inherent in the method must be acknowledged. Firstly, as a pilot study, our investigation involved a small number of participants from a single company, and our findings are not statistically generalizable to a broader population. Instead, we anticipate that researchers and practitioners can draw insights from our discussions, learn about our findings, and transfer the knowledge acquired from our pilot case study to their unique situations and contexts.

Additionally, the study focused primarily on participants' perceptions and experiences without comprehensive quantitative metrics to assess the impact of generative AI tools on productivity. Therefore, as with any qualitative research, there is a potential for researcher bias in data interpretation and analysis. To mitigate this threat to validity, we heavily relied on the raw reports provided by the participants, consistently comparing our interpretations with their views throughout our analysis.

Finally, the pilot nature of the study constrained the depth of data collection and analysis, preventing a thorough exploration of the topic. These limitations underscore the need for future research endeavors with broader and more varied samples, integrating both qualitative and quantitative approaches to offer a comprehensive understanding of the topic.

\section{Conclusions} 
\label{sec:conclusions}

In this paper, we have explored the integration of generative AI tools into software development workflows, aiming to understand their impact on productivity from the perspective of software professionals. Our goal was to provide an understanding of how these tools can be utilized within real-world projects. Through a pilot case study involving software professionals, we collected insights into their experiences while integrating these tools into their daily work routines.

Our findings revealed a generally positive perception of generative AI tools among participants. These tools were particularly valued for their ability to streamline workflows through learning opportunities, optimize time, and facilitate the creation of relevant and insightful content. However, the practitioners reported challenges that mainly revolved around reliability concerns and difficulties in obtaining the desired outcomes, forcing them to manually fix the obtained outcomes for inaccuracies or inconsistencies in generated content.

In conclusion, our pilot case study provides insights into the integration of generative AI tools within software development practices. While our findings suggest promising benefits associated with the utilization of these tools, it is important to address the identified limitations through future research efforts to seamlessly integrate them into the software development process. Overall, our study sets the stage for continued exploration into the evolving landscape of software development practices with the integration of generative AI tools.

\bibliographystyle{ACM-Reference-Format}
\bibliography{bib}


\begin{thebibliography}{36}


\ifx \showCODEN    \undefined \def \showCODEN     #1{\unskip}     \fi
\ifx \showDOI      \undefined \def \showDOI       #1{#1}\fi
\ifx \showISBNx    \undefined \def \showISBNx     #1{\unskip}     \fi
\ifx \showISBNxiii \undefined \def \showISBNxiii  #1{\unskip}     \fi
\ifx \showISSN     \undefined \def \showISSN      #1{\unskip}     \fi
\ifx \showLCCN     \undefined \def \showLCCN      #1{\unskip}     \fi
\ifx \shownote     \undefined \def \shownote      #1{#1}          \fi
\ifx \showarticletitle \undefined \def \showarticletitle #1{#1}   \fi
\ifx \showURL      \undefined \def \showURL       {\relax}        \fi
\providecommand\bibfield[2]{#2}
\providecommand\bibinfo[2]{#2}
\providecommand\natexlab[1]{#1}
\providecommand\showeprint[2][]{arXiv:#2}

\bibitem[Albrecht(1979)]%
        {albrecht1979}
\bibfield{author}{\bibinfo{person}{A.~J. Albrecht}.} \bibinfo{year}{1979}\natexlab{}.
\newblock \showarticletitle{Measuring Application Development Productivity}. In \bibinfo{booktitle}{\emph{Proceedings of IBM Applications Development Symposium}}. \bibinfo{address}{Monterey}, \bibinfo{pages}{83}.
\newblock


\bibitem[{Amazon Web Services}(2023)]%
        {aws_codewhisperer}
\bibfield{author}{\bibinfo{person}{{Amazon Web Services}}.} \bibinfo{year}{2023}\natexlab{}.
\newblock \bibinfo{title}{Amazon CodeWhisperer}.
\newblock \bibinfo{howpublished}{\url{https://aws.amazon.com/codewhisperer/}}.
\newblock
\newblock
\shownote{Accessed: 2023-12-10}.


\bibitem[Boehm et~al\mbox{.}(2000)]%
        {boehm2000}
\bibfield{author}{\bibinfo{person}{Barry Boehm} {et~al\mbox{.}}} \bibinfo{year}{2000}\natexlab{}.
\newblock \bibinfo{booktitle}{\emph{Software Cost Estimation with COCOMO II}}.
\newblock \bibinfo{publisher}{Prentice Hall}, \bibinfo{address}{Upper Saddle River}.
\newblock


\bibitem[Cambon et~al\mbox{.}(2023)]%
        {cambon2023early}
\bibfield{author}{\bibinfo{person}{Alexia Cambon}, \bibinfo{person}{Brent Hecht}, \bibinfo{person}{Benjamin Edelman}, \bibinfo{person}{Donald Ngwe}, \bibinfo{person}{Sonia Jaffe}, \bibinfo{person}{Amy Heger}, \bibinfo{person}{Mihaela Vorvoreanu}, \bibinfo{person}{Sida Peng}, \bibinfo{person}{Jake Hofman}, \bibinfo{person}{Alex Farach}, {et~al\mbox{.}}} \bibinfo{year}{2023}\natexlab{}.
\newblock \bibinfo{booktitle}{\emph{Early LLM-based Tools for Enterprise Information Workers Likely Provide Meaningful Boosts to Productivity}}.
\newblock \bibinfo{type}{{T}echnical {R}eport}. \bibinfo{institution}{MSFT Technical Report. https://www. microsoft. com/en-us/research~…}.
\newblock


\bibitem[Charmaz(2014)]%
        {charmaz2014constructing}
\bibfield{author}{\bibinfo{person}{Kathy Charmaz}.} \bibinfo{year}{2014}\natexlab{}.
\newblock \bibinfo{booktitle}{\emph{Constructing grounded theory}}.
\newblock \bibinfo{publisher}{sage}.
\newblock


\bibitem[Cruzes and Dyba(2011)]%
        {cruzes2011recommended}
\bibfield{author}{\bibinfo{person}{Daniela~S Cruzes} {and} \bibinfo{person}{Tore Dyba}.} \bibinfo{year}{2011}\natexlab{}.
\newblock \showarticletitle{Recommended steps for thematic synthesis in software engineering}. In \bibinfo{booktitle}{\emph{2011 international symposium on empirical software engineering and measurement}}. IEEE, \bibinfo{pages}{275--284}.
\newblock


\bibitem[Digital(2023)]%
        {mckinsey_generative_ai}
\bibfield{author}{\bibinfo{person}{McKinsey Digital}.} \bibinfo{year}{2023}\natexlab{}.
\newblock \bibinfo{title}{Unleashing developer productivity with generative AI}.
\newblock \bibinfo{howpublished}{\url{https://www.mckinsey.com/capabilities/mckinsey-digital/our-insights/unleashing-developer-productivity-with-generative-ai}}.
\newblock
\newblock
\shownote{Acessed in Mar 22, 2024}.


\bibitem[Ebert and Louridas(2023)]%
        {ebert2023generative}
\bibfield{author}{\bibinfo{person}{Christof Ebert} {and} \bibinfo{person}{Panos Louridas}.} \bibinfo{year}{2023}\natexlab{}.
\newblock \showarticletitle{Generative AI for software practitioners}.
\newblock \bibinfo{journal}{\emph{IEEE Software}} \bibinfo{volume}{40}, \bibinfo{number}{4} (\bibinfo{year}{2023}), \bibinfo{pages}{30--38}.
\newblock


\bibitem[Forsgren et~al\mbox{.}(2021)]%
        {forsgren2021space}
\bibfield{author}{\bibinfo{person}{Nicole Forsgren}, \bibinfo{person}{Margaret-Anne Storey}, \bibinfo{person}{Chandra Maddila}, \bibinfo{person}{Thomas Zimmermann}, \bibinfo{person}{Brian Houck}, {and} \bibinfo{person}{Jenna Butler}.} \bibinfo{year}{2021}\natexlab{}.
\newblock \showarticletitle{The SPACE of Developer Productivity: There's more to it than you think.}
\newblock \bibinfo{journal}{\emph{Queue}} \bibinfo{volume}{19}, \bibinfo{number}{1} (\bibinfo{year}{2021}), \bibinfo{pages}{20--48}.
\newblock


\bibitem[Github(2021)]%
        {githubcopilot}
\bibfield{author}{\bibinfo{person}{Github}.} \bibinfo{year}{2021}\natexlab{}.
\newblock \bibinfo{title}{GitHub Copilot}.
\newblock \bibinfo{howpublished}{\url{https://copilot.github.com}}.
\newblock
\newblock
\shownote{Accessed on November 23, 2023}.


\bibitem[Guerrero-Calvache and Hern{\'a}ndez(2022)]%
        {guerrero2022team}
\bibfield{author}{\bibinfo{person}{Marcela Guerrero-Calvache} {and} \bibinfo{person}{Giovanni Hern{\'a}ndez}.} \bibinfo{year}{2022}\natexlab{}.
\newblock \showarticletitle{Team productivity in agile software development: a systematic mapping study}. In \bibinfo{booktitle}{\emph{International Conference on Applied Informatics}}. Springer, \bibinfo{pages}{455--471}.
\newblock


\bibitem[Hoegl et~al\mbox{.}(2003)]%
        {hoegl2003teamwork}
\bibfield{author}{\bibinfo{person}{Martin Hoegl}, \bibinfo{person}{K~Praveen Parboteeah}, {and} \bibinfo{person}{Hans~Georg Gemuenden}.} \bibinfo{year}{2003}\natexlab{}.
\newblock \showarticletitle{When teamwork really matters: task innovativeness as a moderator of the teamwork--performance relationship in software development projects}.
\newblock \bibinfo{journal}{\emph{Journal of Engineering and Technology Management}} \bibinfo{volume}{20}, \bibinfo{number}{4} (\bibinfo{year}{2003}), \bibinfo{pages}{281--302}.
\newblock


\bibitem[Kalliamvakou(2022)]%
        {github_copilot_impact}
\bibfield{author}{\bibinfo{person}{Eirini Kalliamvakou}.} \bibinfo{year}{2022}\natexlab{}.
\newblock \bibinfo{title}{Research: quantifying GitHub Copilot's impact on developer productivity and happiness}.
\newblock \bibinfo{howpublished}{\url{https://github.blog/2022-09-07-research-quantifying-github-copilots-impact-on-developer-productivity-and-happiness/}}.
\newblock
\newblock
\shownote{Acessed in Mar 22, 2024}.


\bibitem[Kim et~al\mbox{.}(2019)]%
        {kim2019understanding}
\bibfield{author}{\bibinfo{person}{Young-Ho Kim}, \bibinfo{person}{Eun~Kyoung Choe}, \bibinfo{person}{Bongshin Lee}, {and} \bibinfo{person}{Jinwook Seo}.} \bibinfo{year}{2019}\natexlab{}.
\newblock \showarticletitle{Understanding personal productivity: How knowledge workers define, evaluate, and reflect on their productivity}. In \bibinfo{booktitle}{\emph{Proceedings of the 2019 CHI Conference on Human Factors in Computing Systems}}. \bibinfo{pages}{1--12}.
\newblock


\bibitem[Lakhanpal(1993)]%
        {lakhanpal1993}
\bibfield{author}{\bibinfo{person}{B Lakhanpal}.} \bibinfo{year}{1993}\natexlab{}.
\newblock \showarticletitle{Understanding the factors influencing the performance of software development groups: An exploratory group-level analysis}.
\newblock \bibinfo{journal}{\emph{Information and Software Technology}} \bibinfo{volume}{35}, \bibinfo{number}{8} (\bibinfo{year}{1993}), \bibinfo{pages}{468--473}.
\newblock
\showISSN{0950-5849}
\urldef\tempurl%
\url{https://doi.org/10.1016/0950-5849(93)90044-4}
\showDOI{\tempurl}


\bibitem[Li et~al\mbox{.}(2024)]%
        {li2024sheetcopilot}
\bibfield{author}{\bibinfo{person}{Hongxin Li}, \bibinfo{person}{Jingran Su}, \bibinfo{person}{Yuntao Chen}, \bibinfo{person}{Qing Li}, {and} \bibinfo{person}{ZHAO-XIANG ZHANG}.} \bibinfo{year}{2024}\natexlab{}.
\newblock \showarticletitle{SheetCopilot: Bringing Software Productivity to the Next Level through Large Language Models}.
\newblock \bibinfo{journal}{\emph{Advances in Neural Information Processing Systems}}  \bibinfo{volume}{36} (\bibinfo{year}{2024}).
\newblock


\bibitem[Lindsj{\o}rn et~al\mbox{.}(2016)]%
        {lindsjorn2016teamwork}
\bibfield{author}{\bibinfo{person}{Yngve Lindsj{\o}rn}, \bibinfo{person}{Dag~IK Sj{\o}berg}, \bibinfo{person}{Torgeir Dings{\o}yr}, \bibinfo{person}{Gunnar~R Bergersen}, {and} \bibinfo{person}{Tore Dyb{\aa}}.} \bibinfo{year}{2016}\natexlab{}.
\newblock \showarticletitle{Teamwork quality and project success in software development: A survey of agile development teams}.
\newblock \bibinfo{journal}{\emph{Journal of Systems and Software}}  \bibinfo{volume}{122} (\bibinfo{year}{2016}), \bibinfo{pages}{274--286}.
\newblock


\bibitem[Moll{\'e}ri et~al\mbox{.}(2016)]%
        {molleri2016survey}
\bibfield{author}{\bibinfo{person}{Jefferson~Seide Moll{\'e}ri}, \bibinfo{person}{Kai Petersen}, {and} \bibinfo{person}{Emilia Mendes}.} \bibinfo{year}{2016}\natexlab{}.
\newblock \showarticletitle{Survey guidelines in software engineering: An annotated review}. In \bibinfo{booktitle}{\emph{Proceedings of the 10th ACM/IEEE international symposium on empirical software engineering and measurement}}. \bibinfo{pages}{1--6}.
\newblock


\bibitem[Monteiro et~al\mbox{.}(2016)]%
        {monteiro2016innovative}
\bibfield{author}{\bibinfo{person}{Cleviton~VF Monteiro}, \bibinfo{person}{Fabio~QB da Silva}, {and} \bibinfo{person}{Luiz~Fernando Capretz}.} \bibinfo{year}{2016}\natexlab{}.
\newblock \showarticletitle{The innovative behaviour of software engineers: Findings from a pilot case study}. In \bibinfo{booktitle}{\emph{Proceedings of the 10th ACM/IEEE International Symposium on Empirical Software Engineering and Measurement}}. \bibinfo{pages}{1--10}.
\newblock


\bibitem[Monteiro et~al\mbox{.}(2023)]%
        {monteiro2023end}
\bibfield{author}{\bibinfo{person}{Mauricio Monteiro}, \bibinfo{person}{Bruno~Castelo Branco}, \bibinfo{person}{Samuel Silvestre}, \bibinfo{person}{Guilherme Avelino}, {and} \bibinfo{person}{Marco~Tulio Valente}.} \bibinfo{year}{2023}\natexlab{}.
\newblock \showarticletitle{End-to-End Software Construction using ChatGPT: An Experience Report}.
\newblock \bibinfo{journal}{\emph{arXiv preprint arXiv:2310.14843}} (\bibinfo{year}{2023}).
\newblock


\bibitem[Nam et~al\mbox{.}(2024)]%
        {nam2024using}
\bibfield{author}{\bibinfo{person}{Daye Nam}, \bibinfo{person}{Andrew Macvean}, \bibinfo{person}{Vincent Hellendoorn}, \bibinfo{person}{Bogdan Vasilescu}, {and} \bibinfo{person}{Brad Myers}.} \bibinfo{year}{2024}\natexlab{}.
\newblock \showarticletitle{Using an llm to help with code understanding}. In \bibinfo{booktitle}{\emph{2024 IEEE/ACM 46th International Conference on Software Engineering (ICSE)}}. IEEE Computer Society, \bibinfo{pages}{881--881}.
\newblock


\bibitem[Noy and Zhang(2023)]%
        {noy2023experimental}
\bibfield{author}{\bibinfo{person}{Shakked Noy} {and} \bibinfo{person}{Whitney Zhang}.} \bibinfo{year}{2023}\natexlab{}.
\newblock \showarticletitle{Experimental evidence on the productivity effects of generative artificial intelligence}.
\newblock \bibinfo{journal}{\emph{Science}} \bibinfo{volume}{381}, \bibinfo{number}{6654} (\bibinfo{year}{2023}), \bibinfo{pages}{187--192}.
\newblock


\bibitem[OpenAI(2023)]%
        {OpenAIChatGPT}
\bibfield{author}{\bibinfo{person}{OpenAI}.} \bibinfo{year}{2023}\natexlab{}.
\newblock \bibinfo{title}{ChatGPT: Optimizing Language Models for Dialogue}.
\newblock \bibinfo{howpublished}{\url{https://openai.com/blog/chatgpt}}.
\newblock
\newblock
\shownote{Acessado em 10 de dezembro de 2023}.


\bibitem[Peng et~al\mbox{.}(2023)]%
        {peng2023impact}
\bibfield{author}{\bibinfo{person}{Sida Peng}, \bibinfo{person}{Eirini Kalliamvakou}, \bibinfo{person}{Peter Cihon}, {and} \bibinfo{person}{Mert Demirer}.} \bibinfo{year}{2023}\natexlab{}.
\newblock \showarticletitle{The impact of ai on developer productivity: Evidence from github copilot}.
\newblock \bibinfo{journal}{\emph{arXiv preprint arXiv:2302.06590}} (\bibinfo{year}{2023}).
\newblock


\bibitem[Prikladnicki et~al\mbox{.}(2013)]%
        {prikladnicki2013cooperative}
\bibfield{author}{\bibinfo{person}{Rafael Prikladnicki}, \bibinfo{person}{Yvonne Dittrich}, \bibinfo{person}{Helen Sharp}, \bibinfo{person}{Cleidson De~Souza}, \bibinfo{person}{Marcelo Cataldo}, {and} \bibinfo{person}{Rashina Hoda}.} \bibinfo{year}{2013}\natexlab{}.
\newblock \showarticletitle{Cooperative and human aspects of software engineering: Chase 2013}.
\newblock \bibinfo{journal}{\emph{ACM SIGSOFT Software Engineering Notes}} \bibinfo{volume}{38}, \bibinfo{number}{5} (\bibinfo{year}{2013}), \bibinfo{pages}{34--37}.
\newblock


\bibitem[Rodr{\'\i}guez et~al\mbox{.}(2012)]%
        {rodriguez2012empirical}
\bibfield{author}{\bibinfo{person}{Daniel Rodr{\'\i}guez}, \bibinfo{person}{MA Sicilia}, \bibinfo{person}{E Garc{\'\i}a}, {and} \bibinfo{person}{Rachel Harrison}.} \bibinfo{year}{2012}\natexlab{}.
\newblock \showarticletitle{Empirical findings on team size and productivity in software development}.
\newblock \bibinfo{journal}{\emph{Journal of Systems and Software}} \bibinfo{volume}{85}, \bibinfo{number}{3} (\bibinfo{year}{2012}), \bibinfo{pages}{562--570}.
\newblock


\bibitem[Runeson and H{\"o}st(2009)]%
        {runeson2009guidelines}
\bibfield{author}{\bibinfo{person}{Per Runeson} {and} \bibinfo{person}{Martin H{\"o}st}.} \bibinfo{year}{2009}\natexlab{}.
\newblock \showarticletitle{Guidelines for conducting and reporting case study research in software engineering}.
\newblock \bibinfo{journal}{\emph{Empirical software engineering}}  \bibinfo{volume}{14} (\bibinfo{year}{2009}), \bibinfo{pages}{131--164}.
\newblock


\bibitem[Ruvimova et~al\mbox{.}(2022)]%
        {ruvimova2022exploratory}
\bibfield{author}{\bibinfo{person}{Anastasia Ruvimova}, \bibinfo{person}{Alexander Lill}, \bibinfo{person}{Jan Gugler}, \bibinfo{person}{Lauren Howe}, \bibinfo{person}{Elaine Huang}, \bibinfo{person}{Gail Murphy}, {and} \bibinfo{person}{Thomas Fritz}.} \bibinfo{year}{2022}\natexlab{}.
\newblock \showarticletitle{An exploratory study of productivity perceptions in software teams}. In \bibinfo{booktitle}{\emph{Proceedings of the 44th International Conference on Software Engineering}}. \bibinfo{pages}{99--111}.
\newblock


\bibitem[Sadowski and Zimmermann(2019)]%
        {sadowski2019rethinking}
\bibfield{author}{\bibinfo{person}{Caitlin Sadowski} {and} \bibinfo{person}{Thomas Zimmermann}.} \bibinfo{year}{2019}\natexlab{}.
\newblock \bibinfo{booktitle}{\emph{Rethinking productivity in software engineering}}.
\newblock \bibinfo{publisher}{Springer Nature}.
\newblock


\bibitem[Seaman(1999)]%
        {seaman1999qualitative}
\bibfield{author}{\bibinfo{person}{Carolyn~B. Seaman}.} \bibinfo{year}{1999}\natexlab{}.
\newblock \showarticletitle{Qualitative methods in empirical studies of software engineering}.
\newblock \bibinfo{journal}{\emph{IEEE Transactions on software engineering}} \bibinfo{volume}{25}, \bibinfo{number}{4} (\bibinfo{year}{1999}), \bibinfo{pages}{557--572}.
\newblock


\bibitem[Sikand et~al\mbox{.}(2024)]%
        {sikand2024much}
\bibfield{author}{\bibinfo{person}{Samarth Sikand}, \bibinfo{person}{Kanchanjot~Kaur Phokela}, \bibinfo{person}{Vibhu~Saujanya Sharma}, \bibinfo{person}{Kapil Singi}, \bibinfo{person}{Vikrant Kaulgud}, \bibinfo{person}{Teresa Tung}, \bibinfo{person}{Pragya Sharma}, {and} \bibinfo{person}{Adam~P Burden}.} \bibinfo{year}{2024}\natexlab{}.
\newblock \showarticletitle{How much SPACE do metrics have in GenAI assisted software development?}. In \bibinfo{booktitle}{\emph{Proceedings of the 17th Innovations in Software Engineering Conference}}. \bibinfo{pages}{1--5}.
\newblock


\bibitem[Strode et~al\mbox{.}(2022)]%
        {strode2022teamwork}
\bibfield{author}{\bibinfo{person}{Diane Strode}, \bibinfo{person}{Torgeir Dings{\o}yr}, {and} \bibinfo{person}{Yngve Lindsjorn}.} \bibinfo{year}{2022}\natexlab{}.
\newblock \showarticletitle{A teamwork effectiveness model for agile software development}.
\newblock \bibinfo{journal}{\emph{Empirical Software Engineering}} \bibinfo{volume}{27}, \bibinfo{number}{2} (\bibinfo{year}{2022}), \bibinfo{pages}{56}.
\newblock


\bibitem[Wohlin and Ahlgren(1995)]%
        {wohlin1995soft}
\bibfield{author}{\bibinfo{person}{Claes Wohlin} {and} \bibinfo{person}{Mattias Ahlgren}.} \bibinfo{year}{1995}\natexlab{}.
\newblock \showarticletitle{Soft factors and their impact on time to market}.
\newblock \bibinfo{journal}{\emph{Software Quality Journal}} \bibinfo{volume}{4}, \bibinfo{number}{3} (\bibinfo{year}{1995}), \bibinfo{pages}{189--205}.
\newblock
\urldef\tempurl%
\url{https://doi.org/10.1007/bf01351923}
\showDOI{\tempurl}


\bibitem[Xie and Memon(2008)]%
        {xie2008using}
\bibfield{author}{\bibinfo{person}{Qing Xie} {and} \bibinfo{person}{Atif~M Memon}.} \bibinfo{year}{2008}\natexlab{}.
\newblock \showarticletitle{Using a pilot study to derive a GUI model for automated testing}.
\newblock \bibinfo{journal}{\emph{ACM Transactions on Software Engineering and Methodology (TOSEM)}} \bibinfo{volume}{18}, \bibinfo{number}{2} (\bibinfo{year}{2008}), \bibinfo{pages}{1--35}.
\newblock


\bibitem[Yin(1994)]%
        {yin1994discovering}
\bibfield{author}{\bibinfo{person}{Robert~K Yin}.} \bibinfo{year}{1994}\natexlab{}.
\newblock \showarticletitle{Discovering the future of the case study. Method in evaluation research}.
\newblock \bibinfo{journal}{\emph{Evaluation practice}} \bibinfo{volume}{15}, \bibinfo{number}{3} (\bibinfo{year}{1994}), \bibinfo{pages}{283--290}.
\newblock


\bibitem[Zhang et~al\mbox{.}(2023)]%
        {zhang2023practices}
\bibfield{author}{\bibinfo{person}{Beiqi Zhang}, \bibinfo{person}{Peng Liang}, \bibinfo{person}{Xiyu Zhou}, \bibinfo{person}{Aakash Ahmad}, {and} \bibinfo{person}{Muhammad Waseem}.} \bibinfo{year}{2023}\natexlab{}.
\newblock \showarticletitle{Practices and challenges of using github copilot: An empirical study}.
\newblock \bibinfo{journal}{\emph{arXiv preprint arXiv:2303.08733}} (\bibinfo{year}{2023}).
\newblock


\end{thebibliography}

\end{document}